\pdfoutput=1
\documentclass[fleqn,usenatbib]{mnras}

\usepackage[T1]{fontenc}
\usepackage{newtxtext,newtxmath}
\usepackage{graphicx}
\usepackage{amsmath}
\usepackage{xcolor}

\providecommand{\sovast}{Soviet Ast.}
\providecommand{\aap}{A\&A}
\providecommand{\apj}{ApJ}
\providecommand{\mnras}{MNRAS}

\newcommand{\be}{\begin{equation}}
\newcommand{\ee}{\end{equation}}
\newcommand{\aaa}{\textsf{a}}
\renewcommand{\d}{{\mathrm{d}}}

\defcitealias{Gradshteyn_8}{GR}

\title[Basis pairs for radial perturbations]{Two sets of potential-density basis pairs for the study \\of radial perturbations in collisionless spherical stellar systems}

\author[E. V. Polyachenko and I. G. Shukhman]{
Evgeny~V.~Polyachenko$^{1}$\thanks{E-mail: epolyach@inasan.ru}
and Ilia~G.~Shukhman$^{2}$\thanks{E-mail: shukhman@iszf.irk.ru}
\\
$^{1}$Institute of Astronomy, Russian Academy of Sciences, 48 Pyatnitskaya st, Moscow 119017, Russia\\
$^{2}$Institute of Solar-Terrestrial Physics, Russian Academy of Sciences, Siberian Branch, P.O. Box 291, Irkutsk 664033, Russia
}

\date{Accepted 2026 September 3. Received 2026 September 2; in original form 2026 July 4}
\pubyear{2026}

\begin{document}
\label{firstpage}
\pagerange{\pageref{firstpage}--\pageref{lastpage}}
\maketitle

\begin{abstract}
The Kalnajs matrix method is a widely used framework for
studying the global linear stability and possible Landau damping
of collisionless stellar systems. However, for radial
perturbations ($l=0$) in three-dimensional spherical models with
infinite boundaries, standard biorthogonal sets such as the
Clutton-Brock basis often converge slowly. This stems from a
physical constraint: mass-conserving radial modes force the
perturbed potential to decay faster than a point mass at
infinity, whereas individual Clutton-Brock elements carry a
fictitious net mass and decay only as $\mathcal{O}(1/r)$,
producing unphysical asymptotic tails. We construct two new
families of potential-density basis pairs that are free of these
tails by construction. The first modifies the Clutton-Brock set: a specific
linear combination of adjacent elements analytically cancels the
leading $\mathcal{O}(1/r)$ term, giving potentials that decay as
$\mathcal{O}(1/r^3)$ and densities as $\mathcal{O}(1/r^5)$. This
breaks strict biorthogonality but yields a compact tridiagonal
Gram matrix that enters the response equation at
negligible additional cost. The second family is built from Jacobi
polynomials that embed the required $\mathcal{O}(1/r^2)$ potential
decay directly into their construction while retaining strict
diagonal biorthogonality. Numerical tests demonstrate convergence that is uniform in radius
for both expansions: the truncation error decreases exponentially
when the asymptotic tail of the expanded potential matches the
parity of the basis elements, and algebraically otherwise. Both
bases reduce the dimension of the response matrix required for a
given accuracy and are suited to studying radial perturbations in
open stellar systems.
\end{abstract}


\section{Introduction}
 
The global stability of spherical stellar systems against
radial perturbations is a classical and well-studied
problem in galactic dynamics, with its foundations firmly established by \citet{Ant61,Ant62}. Instead of stability thresholds, current studies of spherical models focus on the mechanism of perturbation damping in stable systems and the possible existence of Landau quasi-modes.
 
Investigating these phenomena relies heavily on the matrix
method, originally formulated by \citet{Kalnajs77} for
stellar disks and subsequently generalized to spherical
configurations by \citet{PS81}. The computational efficiency
and spectral accuracy of the matrix method are critically
dependent on the choice of an optimal set of basis function
pairs for the potential and density expansion \citep[see, e.g.,][]{BPRV94}.
 
A fundamental physical constraint in this context is the
preservation of the total mass of the system. For any
physical radial perturbation that redistributes matter
without altering the total mass, the net perturbed mass
must strictly vanish ($\delta M_{\text{tot}} = 0$). From
potential theory, this zero-mass condition dictates that
the perturbed potential $\delta\Phi(r)$ must decay at
large distances faster than the standard monopole
field, i.e., faster than $\mathcal{O}(1/r)$ as $r \to \infty$.

Despite the clear physical necessity of this requirement,
self-consistent potential-density basis sets that naturally
incorporate the $\delta M_{\text{tot}} = 0$ condition
are generally absent in the literature. Standard expansions,
such as the classical \citet{CB73} or \citet{HO92} bases,
including those listed in the detailed work by \citet{Lil2018},
possess a dominant $\mathcal{O}(1/r)$ potential tail at
infinity, making them poorly suited for expanding
fast-decaying radial modes.
 
In the present work, we address this gap by developing two families of radial potential-density basis pairs designed specifically for radial perturbations in spherical systems. The first family provides an
asymptotic decay of the perturbed potential of
$\mathcal{O}(1/r^3)$ at large distances, while the second
family has a slower decay of $\mathcal{O}(1/r^2)$, which
still satisfies the zero-mass condition. Both formulations enforce the mass-conservation law by construction and avoid the slow convergence of standard expansions in resolving the fine features of resonant damping and quasi-modes.

The first family is a modified Clutton-Brock (CB) basis
optimized for core-concentrated perturbations. It features
an asymptotic potential decay of $\mathcal{O}(1/r^3)$ and
a density decay of $\mathcal{O}(1/r^5)$. Following the
Galerkin recombination idea of \citet{Shen94,Shen95} and
\citet{Boyd2001}, we pair adjacent standard elements to cancel
their $\mathcal{O}(1/r)$ tails. 
This replaces the identity matrix in the Kalnajs eigenvalue problem by a sparse, symmetric tridiagonal Gram matrix, a negligible overhead far outweighed by the improved convergence.

To quantify the quality of a truncated expansion we use a local
asymptotic error measure $F_A(r)$, defined in
Section~\ref{sec:verification} as the deviation of the radially
weighted partial sum from its exact limit at infinity, with the
weight matched to the decay rate of each basis. As an
application, the expansion of a test potential representing the
difference of two Plummer potentials \citep{Plu_1911}, which
decays as $1/r^3$, over the modified basis converges exponentially.
 
The second family is a newly developed Even-Power basis
designed specifically to capture potential perturbations such as,
for example, the dilation modes considered recently by
\citet{PS23,PS24}. Built upon Jacobi polynomials $P^{(3/2,1/2)}_{n-1}(x)$
mapped via the compact variable $x = (1-r^2)/(1+r^2)$, this
basis natively enforces a slow potential decay of
$\mathcal{O}(1/r^2)$ and a density decay of $\mathcal{O}(1/r^4)$.
 
A key mathematical advantage of this new family is its
strict diagonal biorthogonality, which yields an identity
Gram matrix without requiring any algebraic
recombination. Furthermore, we demonstrate that
the plateau level of the error at infinity, $F_A(\infty)$, decreases steadily as the truncation order increases.

The paper is organized as follows. Section~2 constructs the
modified Clutton-Brock basis, derives its tridiagonal Gram
matrix, and tests its convergence on a mass-conserving
Plummer-difference potential. Section~3 develops the Even-Power basis and examines how its convergence rate depends on the asymptotic structure of the potential being expanded. Section~4 summarizes the results.

\section{Modified Clutton-Brock Basis}

The standard biorthogonal Clutton-Brock basis for radial modes ($l=0$) is given, in a compact trigonometric form equivalent to the original Gegenbauer representation, by \citet{PS24}:
\begin{equation}
\Phi^\alpha(r) = -\frac{2}{\sqrt{\pi\, (4\alpha^2 - 1)}} \frac{\sqrt{1+r^2}}{r} \sin\left[2\alpha \arcsin\frac{1}{\sqrt{1+r^2}}\right],
\label{eq:Phi_alpha}
\end{equation}

\begin{equation}
\rho^\alpha(r) = 2\sqrt{\frac{4\alpha^2 - 1}{\pi}} \frac{1}{r\,(1+r^2)^{3/2}} \sin\left[2\alpha \arcsin\frac{1}{\sqrt{1+r^2}}\right],
\end{equation}
where $\alpha=1, 2, 3, ...$.

\medskip

Expanding the full potential expression (\ref{eq:Phi_alpha}) in powers of $1/r$, we find that the leading-order point-mass contribution is given by:
\begin{equation}
 \Phi^\alpha(r) = -\frac{4\alpha}{\sqrt{\pi(4\alpha^2 - 1)}} \frac{1}{r} + \mathcal{O}\left(\frac{1}{r^3}\right).
\end{equation}
Note that due to the symmetry of the Clutton-Brock construction, the potential expansion contains only odd powers of $1/r$.

We define the modified potential basis functions $\tilde{\Phi}^\alpha(r)$ as a linear combination of two adjacent standard elements:
\begin{equation}
\tilde{\Phi}^\alpha(r) = \Phi^\alpha(r) - K_\alpha \Phi^{\alpha+1}(r).
\label{eq:phi-K}
 \end{equation}
The coupling coefficient $K_\alpha$ is fixed by requiring that the leading $\mathcal{O}(1/r)$ coefficients of the two terms cancel.
This condition yields
 \begin{equation}
  K_\alpha = \frac{\alpha}{\alpha+1} \sqrt{\frac{4\,(\alpha+1)^2 - 1}{4\alpha^2 - 1}}.
  \label{eq:K}
   \end{equation}

   With (\ref{eq:K}) substituted into (\ref{eq:phi-K}), the leading $\mathcal{O}(1/r)$ terms cancel exactly and the modified potential obeys \begin{equation} \tilde{\Phi}^\alpha(r) = \mathcal{O}\left(\frac{1}{r^3}\right) \quad \text{as} \quad r \to \infty. \end{equation} The corresponding modified density $\tilde{\rho}^\alpha(r) = \rho^\alpha(r) - K_\alpha \rho^{\alpha+1}(r)$ follows from the spherical Poisson equation $\Delta \tilde{\Phi}^\alpha = \tilde{\rho}^\alpha$ (no $4\pi$ factor, consistent with the normalization $\int \Phi^\alpha \rho^\beta r^2 dr = -\delta^{\alpha\beta}$); since $\Delta(1/r^3) = 6/r^5$, the density decays as $\tilde{\rho}^\alpha(r) = \mathcal{O}(1/r^5)$ and the mass integrals of individual elements converge absolutely.

The two-term combination (\ref{eq:phi-K}) in fact collapses into a single
polynomial. In the algebraic coordinate $\xi = (r^2-1)/(r^2+1)$ the standard
elements (\ref{eq:Phi_alpha}) can be written through  Chebyshev polynomials of the second
kind $U_{\alpha-1}$, 
\[
\Phi^\alpha(r) =-\frac{4}{\sqrt{\pi\,(4\alpha^2-1)}} \frac{U_{\alpha-1}(\xi)}{\sqrt{1+r^2}}
\] \citep[see][Eqs (1.7a) and (1.9)]{CB73}, i.e.\ Jacobi
polynomials with indices $(1/2,1/2)$, and the contiguous relation
\citep[][Eq.~8.961.5, hereafter GR]{Gradshteyn_8} turns the combination
\[U_{\alpha-1}(\xi) - \frac{\alpha}{\alpha+1}\,U_{\alpha}(\xi)\] into
$(1-\xi)\,P^{(3/2,\,1/2)}_{\alpha-1}(\xi)$ up to a constant factor, where $P^{(3/2,1/2)}_{\alpha-1}$ is the Jacobi polynomial. The
prefactor $1-\xi = 2/(1+r^2)$ supplying the extra decay. This yields the closed form
\begin{equation}
\tilde{\Phi}^\alpha(r) = -\frac{2^{\alpha+2}\,\alpha!}{(2\alpha-1)!!\,
\sqrt{\pi\,(4\alpha^2-1)}}\,
\frac{P^{(3/2,\,1/2)}_{\alpha-1}(\xi)}{(1+r^2)^{3/2}}.
\label{eq:jacobi-cb}
\end{equation}

\subsection{Structure of the Gram Matrix}

In the standard formulation of the Kalnajs matrix method, the strict biorthogonality of the potential-density pairs 
ensures that the Gram matrix \citep[see, e.g.][]{HofKen71} of the expansion, that is, the matrix of scalar products of the basis elements, is simply the identity matrix, once the conventional minus sign of the gravitational scalar product is absorbed into the definition.
However, the transition to the modified basis $\tilde{\Phi}^\alpha = \Phi^\alpha - K_\alpha \Phi^{\alpha+1}$ and $\tilde{\rho}^\alpha = \rho^\alpha - K_\alpha \rho^{\alpha+1}$ alters this structure.

For the modified basis, we define the Gram matrix $\Delta^{\alpha\beta}$ as
\begin{equation}
\Delta^{\alpha\beta} =\! -\!\int_0^\infty \!\!\!\tilde{\Phi}^\alpha(r)\, \tilde{\rho}^\beta(r)\, r^2 \, dr=\!\int_0^\infty\!\! \frac{\d\tilde{\Phi}^\alpha(r)}{\d r}\, \frac{\d\tilde{\Phi}^\beta(r)}{\d r}\, r^2\,\d r.
\end{equation}

Substituting the linear combinations and using the original biorthogonality relation $\int \Phi^\alpha \rho^\beta r^2 dr = -\delta^{\alpha\beta}$ gives $\Delta^{\alpha\alpha} = 1 + K_\alpha^2$, $\Delta^{\alpha,\alpha+1} = \Delta^{\alpha+1,\alpha} = -K_\alpha$, and $\Delta^{\alpha\beta} = 0$ for $|\alpha-\beta| > 1$. The Gram matrix is therefore tridiagonal and symmetric:
\begin{equation}
\mathbf{\Delta} =
\begin{pmatrix}
1+K_1^2 & -K_1 & 0 & \dots & 0 \\
-K_1 & 1+K_2^2 & -K_2 & \dots & 0 \\
0 & -K_2 & 1+K_3^2 & \dots & 0 \\
\vdots & \vdots & \vdots & \ddots & \vdots \\
0 & 0 & 0 & \dots & 1+K_A^2
\end{pmatrix}.
\end{equation}
Here $A$ denotes the truncation order of the expansion, i.e.\ the number of retained basis elements, $\alpha = 1, \dots, A$.

In the generalized Kalnajs response framework, this sparse matrix replaces the standard identity matrix. The matrix equation for locating the complex frequencies $\omega$ of the radial modes takes the form:
\begin{equation}
\text{Det}\left|\left| \Delta^{\alpha\beta} - \mathcal{M}^{\alpha\beta}(\omega) \right| \right| = 0,
\end{equation}
where $\mathcal{M}^{\alpha\beta}(\omega)$ is the standard response matrix \citep[see][]{Kalnajs77,PS81}, calculated here in the modified potential basis. Because $\mathbf{\Delta}$ is tridiagonal, its inclusion adds no appreciable cost to the evaluation of the determinant, while the physical regularization at infinity reduces the truncation order $A$ at which $\mathcal{M}^{\alpha\beta}$ attains a given accuracy.

In principle, any basis with the required asymptotic behaviour
can be adopted, for instance one built on Laguerre polynomials on the
half-line \citepalias[Eq.~8.970.1]{Gradshteyn_8}, even at the price of a Gram matrix that is no longer sparse.
This in itself is not an obstacle: the expansion coefficients follow
from a single $A \times A$ linear solve,  the matrix equation above
retains its form with the dense matrix in place of the tridiagonal one,
and the dominant computational load, the phase-space quadratures for
the response matrix $\mathcal{M}^{\alpha\beta}(\omega)$, is insensitive
to the choice of basis. Two practical criteria, however, remain.

The first is the conditioning of $\mathbf{\Delta}$. For a
symmetric positive-definite matrix the condition number
$\mathrm{cond}\,(\mathbf{\Delta})$, the ratio of its largest and
smallest eigenvalues, shows by how much relative errors may be
amplified in solving the linear system for the expansion coefficients \citep[see, e.g.,][]{Higham2002, Press2007}.
For the tridiagonal matrix above it grows only quadratically with $A$,
$\mathrm{cond}\,(\mathbf{\Delta}) \simeq 4A^2/\pi^2$, whereas for a
generic non-biorthogonal set it is not controlled and must be examined
case by case.

The second is the convergence rate, governed by the matching
between the asymptotic behaviour of the basis and that of the expanded
potential (Section~\ref{sec:rate}). In the Laguerre example, one may
build the density elements on the half-line functions, $\rho^n(r)
\propto L^{(2)}_n(r/s)\,{\rm e}^{-r/s}$ with a scale parameter $s$, and
obtain the potential companions from the Poisson equation; since
$\int_0^\infty \d r\,r^2\,\rho^n(r)\propto\int_0^\infty L^{(2)}_n(x)\,x^2 {\rm e}^{-x}\,{\rm d}x = 0$ for $n \geq
1$  \citepalias[due to the orthogonality of the Laguerre polynomials, see, e.g.,][Eq.~7.414.3]{Gradshteyn_8},   every element except the zeroth carries exactly zero net mass, so
the required asymptotic behaviour is indeed built in. The convergence
rate, however, is limited by the mismatch between the exponential
falloff of the basis and the power-law tails of perturbations in
typical models \citep{Boyd2001}; in addition, the scale $s$ enters as a
free parameter requiring adjustment.

\subsection{Numerical Verification of the Modified Basis}
\label{sec:verification}
To test the convergence of the modified basis 
we choose a mass-conserving test potential: the difference
of two equal-mass Plummer potentials with distinct scale radii, $b_1 \neq b_2$:
\begin{equation}
\delta\Phi(r) = -\frac{1}{\sqrt{b_1^2 + r^2}} + \frac{1}{\sqrt{b_2^2 + r^2}}. \end{equation}
By construction $\delta M_{\text{tot}} = 0$ identically; at large distances the leading point-mass $\mathcal{O}(1/r)$ tails cancel, yielding \begin{equation} 
\delta\Phi(r) = \frac{b_1^2 - b_2^2}{2r^3} + \mathcal{O}\left(\frac{1}{r^5}\right). \end{equation}
Thus, the {\it asymptotic amplitude} $R$ of the test potential is:
\begin{equation}
R \equiv \lim_{r\to\infty} [r^3 \delta\Phi(r)] = \frac{1}{2}\,(b_1^2 - b_2^2). \end{equation}
In what follows, we set $b_1 = 0.5$ and $b_2 = 1.5$, which gives $R = -1$. To monitor the truncation error at large radii we introduce the asymptotic error function
\begin{equation} F_A(r) = \big|r^3\, \delta\Phi_A(r) - R\big|, \qquad
\delta\Phi_A(r) = \sum_{\alpha=1}^A C^\alpha \tilde{\Phi}^\alpha(r),
\end{equation}
where $\delta\Phi_A$ is the truncated expansion of the test potential, $A$ is the truncation order and $C^\alpha$ are the expansion coefficients, obtained by projecting the test potential onto the modified density elements and solving the linear system with the tridiagonal Gram matrix $\mathbf{\Delta}$. Where the two bases need to be distinguished, we write $F_A^{\rm std}$ and $F_A^{\rm mod}$ for the errors of the standard and modified expansions, respectively. 
Since any finite standard sum retains an uncancelled $\mathcal{O}(1/r)$ tail, $F_A^{\rm std}(r)$ diverges as $\sim r^2$ at large radii. Each modified element, in contrast, decays as $\mathcal{O}(1/r^3)$, so $F_A^{\rm mod}(r)$ remains bounded and levels off at large radii to a constant value, $F_A^{\rm mod}(\infty)$, which decreases as $A$ grows.

\begin{figure}
\centering
\includegraphics[width=\columnwidth]{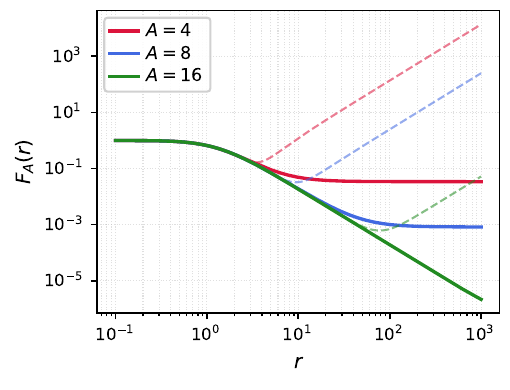}
\caption{The local asymptotic error
$F_A(r) \equiv |r^3 \delta\Phi_A(r) - R|$ for the mass-conserving
Plummer-difference potential ($b_1 = 0.5$, $b_2 = 1.5$, asymptotic
amplitude $R = -1$). Faint dashed lines: the standard
Clutton-Brock expansion, whose unphysical $\mathcal{O}(1/r)$
point-mass tails make the error diverge as $\sim r^2$ at large
radii; bold solid lines: the modified basis, which enforces
$\mathcal{O}(1/r^3)$ decay and stays strictly bounded, with an outer
plateau whose level drops exponentially with $A$. Colours denote
$A = 4, 8, 16$.}
\label{fig:error_plummer}
\end{figure}

The convergence rate is set by how fast the plateau level
$F_A^{\rm mod}(\infty)$ tends to zero. For the
Plummer-difference potential our numerical computations give an
exponential law with decrement $\sigma$,
\begin{equation}
F_A^{\rm mod}(\infty) \propto \exp(-\sigma A),
\quad \sigma \approx 1.099.
\end{equation}
The origin of this value, $\sigma = \ln 3$, is explained by the
singularity analysis in Section~\ref{sec:rate}.
The corresponding error curves are shown in Fig.~\ref{fig:error_plummer}.

\section{Even-Power Biorthogonal Basis}

\subsection{Explicit Form of Potential and Density Functions}
 
To model self-similar configurations such as dilation modes,
which exhibit an intermediate potential fall-off at large
distances while conserving mass ($\delta M_{\text{tot}} = 0$),
we introduce a second biorthogonal radial basis for $l=0$. Its
elements contain only even inverse powers of the radius, so that
as $r \to \infty$:
\begin{equation}
\Phi^n(r) = \frac{C_n}{r^2} + \mathcal{O}\left(\frac{1}{r^4}\right),
\label{eq:C_n}
\end{equation}
\begin{equation}
\rho^n(r) = \frac{D_n}{r^4} + \mathcal{O}\left(\frac{1}{r^6}\right),
\label{eq:D_n}
\end{equation}
where $C_n$ and $D_n$ are constant amplitudes, given in closed
form at the end of this subsection.

For each order $n = 1, 2, 3, \dots$, we define the compact
algebraic coordinate $x \in [-1, 1]$ as 
\be
    x = \frac{1-r^2}{1+r^2}.
    \label{eq:alg-ccord}
\ee
To ensure a strictly diagonal (identity) Gram matrix, the
biorthogonal pair $\{\Phi^n(r), \rho^n(r)\}$ must be constructed
using Jacobi polynomials $P_{n-1}^{(3/2,\,1/2)}(x)$
\citepalias[Eq.~8.960]{Gradshteyn_8}. In closed analytical form,
the potential is given by the normalized integral of the Jacobi polynomial:
\begin{equation}
\Phi^n(r) = - N_n \int_{-1}^{x} P_{n-1}^{(3/2, 1/2)}(x') \, dx',
\end{equation}
where the self-consistent normalization constant $N_n$ reads:
\begin{equation}
N_n = \frac{2^{n-1} \sqrt{(n-1)! \, (n+1)!}}{(2n-1)!! \, \sqrt{\pi}}.
\end{equation}
The companion density elements follow from the spherical
Laplacian, $\rho^n(r) = \Delta \Phi^n(r)$, and the Gram
matrix is the identity:
\begin{equation}
-\int_0^\infty \Phi^n(r) \, \rho^m(r) \, r^2 \, dr = \delta^{nm}.
\end{equation}

The lowest-order analytical functions for the potential and 
companion density profiles are evaluated as:
\begin{equation}
\Phi^1(r) = -\sqrt{\frac{8}{\pi}} \frac{1}{1+r^2},
\end{equation}
\begin{equation}
\Phi^2(r) = -\sqrt{\frac{8}{3\pi}} \frac{1-3r^2}{(1+r^2)^2},
\end{equation}
\begin{equation}
\Phi^3(r) = -\sqrt{\frac{16}{3\pi}} \frac{1-4r^2+3r^4}{(1+r^2)^3},
\end{equation}
\begin{equation}
\rho^1(r) = \sqrt{\frac{32}{\pi}} \frac{3-r^2}{(1+r^2)^3},
\end{equation}
\begin{equation}
\rho^2(r) = \sqrt{\frac{96}{\pi}} \frac{5-10r^2+r^4}{(1+r^2)^4},
\end{equation}
\begin{equation}
\rho^3(r) = \sqrt{\frac{192}{\pi}} \frac{7-35r^2+21r^4-r^6}{(1+r^2)^5}.
\end{equation}
Higher orders follow from the standard Jacobi three-term recurrence \citepalias[Eq.~8.961.2]{Gradshteyn_8}.

The amplitudes in equations (\ref{eq:C_n}) and (\ref{eq:D_n})
are given by:
\begin{equation}
C_n = (-1)^{n} \cdot 2 \sqrt{\frac{n(n+1)}{\pi}}, \quad 
D_n = 2 C_n.
\end{equation}

\subsection{Numerical Verification of the Even-Power Basis}

To demonstrate the advantage of the Even-Power basis over the Clutton-Brock bases,
we evaluate the analogous truncation error, $\delta\Phi_A$ now
denoting the truncated expansion over the Even-Power elements:
\begin{equation}
F_A(r) \equiv \left| r^2 \delta\Phi_A(r) - \lim_{r\to\infty}
\left[ r^2\,\delta\Phi(r) \right] \right|,
\end{equation}
for the dilation mode in the isochrone model \citep[][Eq.~4.54]{Hen_60,BT2008},
recently considered by \citet{PS23,PS24}. For this specific mode,
the exact potential perturbation is given by:
\begin{equation}
\delta\Phi(r) = \frac{1}{b\,a(r)\,[b+a(r)]}, \quad
a(r) = \sqrt{b^2+r^2},
\label{eq:dilat}
\end{equation}
where $b$ is a characteristic scale length. The potential decays as $\mathcal{O}(1/r^2)$ at large distances, and its asymptotic amplitude, now defined with the $r^2$ weight,
is $R \equiv \lim_{r\to\infty} \left[ r^2\, \delta\Phi(r) \right] = 1/b$.

Because of this slow decay, the modified Clutton-Brock basis, 
as well as the standard one, is unable to approximate this 
profile efficiently at large distances.
 
\medskip

Numerical tests with $b = 1$ show that any finite truncation of
the standard Clutton-Brock series fails to cancel the unphysical
monopole tails: evaluated with the same $r^2$ weight, its error
reaches $F_8^{\rm std}(1000) \approx 74$ and keeps growing with
$r$.  Suppressing it below the per cent level at the same distance would require a prohibitively large truncation order.

\begin{figure}
\centering
\includegraphics[width=\columnwidth]{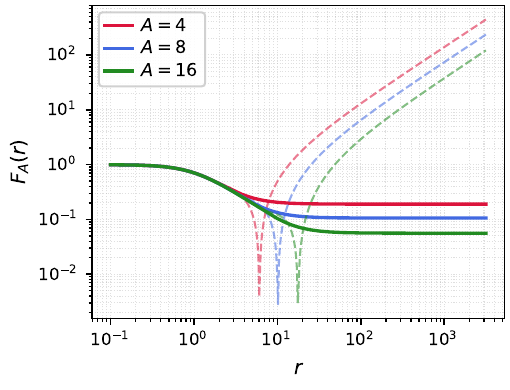}
\caption{The local asymptotic error
$F_A(r) \equiv |r^2 \delta\Phi_A(r)-R|$ for the isochrone
dilation mode ($b=1$, asymptotic amplitude $R = 1$). Faint dashed lines:
the standard Clutton-Brock expansion, whose uncancelled $\mathcal{O}(1/r)$ tails make the error diverge as $\sim r$ at large radii;
the sharp downward spikes mark isolated radii where $r^2\,\delta\Phi_A(r)-R$ changes sign. Bold solid lines: the Even-Power basis, which 
constrains the error to a flat outer plateau whose level decreases algebraically as $1/A$.
Colours denote $A = 4, 8, 16$.}
\label{fig:error_dilation}
\end{figure}

The Even-Power basis, by contrast, keeps the error at large radii
bounded ($F_A \le 1.0$) even at the lowest orders, forming a flat
plateau, as illustrated in Fig.~\ref{fig:error_dilation}.
The convergence is again set by how fast the plateau level $F_A(\infty)$ tends to zero;
for the dilation mode the decay is algebraic,
\begin{equation}
F_A(\infty) \propto \frac{1}{A},
\end{equation}
as shown in Fig.~\ref{fig:scaling} below.

\medskip

\subsection{Dependence of the Convergence Rate on the Expanded Potential}
\label{sec:rate}

The algebraic law $F_A(\infty) \propto 1/A$ found in the previous
subsection is \emph{not} an intrinsic limitation of the Even-Power
basis; it is specific to the analytic structure of the isochrone
dilation mode. Expanding that mode at large radius gives
\begin{equation}
\delta\Phi(r) = \frac{1}{b\,r^2} - \frac{1}{r^3}
+ \mathcal{O}\!\left(\frac{1}{r^5}\right),
\label{eq:dil_tail}
\end{equation}
where the odd term $-1/r^3$ originates from the factor $b+\sqrt{b^2+r^2}$ in
the denominator of (\ref{eq:dilat}). It is independent of $b$ and is responsible for the loss of parity symmetry.
The
Even-Power potentials contain only even inverse powers
($\Phi^n = C_n/r^2 + \mathcal{O}(1/r^4)$), so this odd term cannot be
represented: under the map \eqref{eq:alg-ccord} it is a branch point
$\propto(1+x)^{3/2}$ sitting exactly on the domain boundary $x=-1$,
which caps the convergence at the algebraic rate.

To determine the convergence rate of the basis itself, we
consider a smooth, mass-conserving test potential whose large-$r$ expansion contains only even inverse powers of $r$, with $b$ again denoting a core scale:
\begin{equation}
\delta\Phi(r) = \frac{1}{b^2 + r^2}, \qquad
R \equiv \lim_{r\to\infty} r^2\,\delta\Phi(r) = 1 .
\label{eq:clean_target}
\end{equation}
As a rational function of $x$, i.e.
\[\delta\Phi(x)=\frac{1}{b^2+(1-x)/(1+x)},\]
this profile is analytic on the whole
segment $[-1,1]$ (including the boundary), its \emph{only} singularity
being a simple pole on the real axis of the mapped complex plane $z\in\mathbb{C}$:
\begin{equation}
z_{\text{sing}} = -\frac{b^2 + 1}{b^2 - 1}.
\end{equation}

According to classical results of approximation theory, the asymptotic
convergence rate of an expansion in orthogonal polynomials is
governed by the largest singularity-free Bernstein ellipse
with foci $\pm1$,
$|z-1|+|z+1|=2\aaa=\lambda+\lambda^{-1}$. Here $\lambda>1$ is the sum of the
semi-major axis $\aaa$ and the semi-minor axis $\sqrt{\aaa^2-1}$ of the ellipse \citep[see][and chapter~8 of
\citealt{Trefethen2013} for a modern exposition]{Bernstein1912}.

The ellipse through $z_{\text{sing}}$
has parameter $\lambda=|(b+1)/(b-1)|$, giving exponential decay of
the outer error plateau, $F_A(\infty)\propto\lambda^{-A}$, with decrement
\begin{equation}
\sigma(b) = \ln\lambda = \ln \left| \frac{b + 1}{b - 1} \right|.
\end{equation}
The divergence of $\sigma(b)$ as $b \to 1$ has a simple
interpretation: for $b = 1$ the test potential coincides, up to
normalization, with the first basis element,
$\delta\Phi(r) = 1/(1+r^2) = -\sqrt{\pi/8}\, \Phi^1(r)$, so a
single term represents it exactly and $F_A(\infty)$ vanishes for
all $A \ge 1$; in the mapped plane, the pole $z_{\text{sing}}$
recedes to infinity. We have checked numerically that the
projections onto all higher elements vanish to machine precision.

\begin{figure}
\centering
\includegraphics[width=\columnwidth]{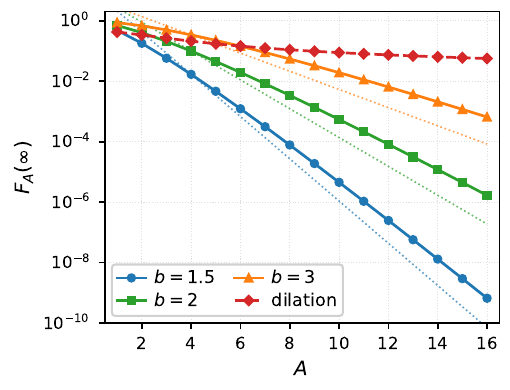}
\caption{Plateau level of the error, $F_A(\infty)$, versus truncation order $A$ for
the Even-Power basis. For smooth potentials with only even
inverse powers of $r$ \eqref{eq:clean_target} the convergence is exponential with a tunable
decrement $\sigma(b)=\ln[(b{+}1)/(b{-}1)]$ (solid; dotted lines show
$\lambda^{-A}$ with $\lambda=(b{+}1)/(b{-}1)=5,3,2$ for $b=1.5,2,3$). The
isochrone dilation mode (red, dashed; $b=2$), whose $\mathcal{O}(1/r^3)$
tail \eqref{eq:dil_tail} cannot be represented by the even-power elements,
is limited to the algebraic rate $\propto 1/A$.}
\label{fig:scaling}
\end{figure}

Direct numerical computation of the expansion coefficients
confirms this law
(Fig.~\ref{fig:scaling}). For $b=2$ successive coefficients
decrease asymptotically by the factor $\lambda=3$; the plateau level
falls as $F_A(\infty)\approx 1.0\times10^{-1},\,3.4\times10^{-3},\,
8.2\times10^{-5},\,1.7\times10^{-6}$ at $A=4,8,12,16$, with a local
decrement approaching the predicted $\sigma=\ln 3 \approx 1.099$. The rate is fully
tunable through the core scale: $b=1.5$ yields $\sigma=\ln 5\approx1.609$
and $b=3$ yields $\sigma=\ln 2\approx0.693$, in quantitative agreement
with the analytic prediction over eight orders of magnitude.

The same argument explains the decrement found in
Section~\ref{sec:verification} for the modified Clutton-Brock
basis: up to a common factor $\sqrt{1+x}$, its elements are
polynomials in the variable \eqref{eq:alg-ccord}, so the
convergence rate is again set by the singularities of the expanded
potential in the mapped plane. The Plummer-difference potential is
singular at $r^2 = -b_{1,2}^2$, that is, at
$x = (1+b_{1,2}^2)/(1-b_{1,2}^2)$; the corresponding Bernstein
ellipses have $\lambda = 3$ for $b_1 = 0.5$ and $\lambda = 5$ for
$b_2 = 1.5$. The smaller value limits the rate, giving
$\sigma = \ln 3 \approx 1.099$, precisely the decrement measured
in Section~\ref{sec:verification}. Its agreement with the $b = 2$
value above is a coincidence of the chosen parameters.

We stress that the exponential rate applies only to potentials
whose asymptotic expansions contain no odd inverse powers of $r$.
Modes such as the isochrone dilation
\eqref{eq:dil_tail}, whose $\mathcal{O}(1/r^3)$ term cannot be
represented by even powers, retain the algebraic convergence of
the previous subsection unless the expansion is augmented with
odd-parity elements, for example standard Clutton-Brock functions,
to absorb that tail.

\section{Conclusion}
 
We have constructed two families of potential-density basis
pairs for the linear analysis of radial ($l=0$) perturbations in
collisionless spherical stellar systems. Both are built around
the mass-conservation constraint $\delta M_{\text{tot}} = 0$,
which individual elements of traditional biorthogonal expansions,
such as the Clutton-Brock and Hernquist-Ostriker bases, violate:
each element carries a fictitious net mass with an
$\mathcal{O}(1/r)$ potential tail that slows the convergence of
the Kalnajs matrix method.

The two families offer complementary advantages:
\begin{enumerate}
\item The modified Clutton-Brock basis, suited to
core-concentrated perturbations, pairs adjacent standard elements
through the analytically derived coefficient $K_\alpha$, following
the Galerkin recombination idea of \citet{Shen94,Shen95} and
\citet{Boyd2001}. The pairing enforces $\mathcal{O}(1/r^3)$
potential and $\mathcal{O}(1/r^5)$ density decay; strict
biorthogonality is replaced by a symmetric tridiagonal Gram matrix
$\mathbf{\Delta}$ that enters the generalized eigenvalue problem
at negligible additional cost.
\item The Even-Power basis, built from Jacobi polynomials in the
mapped coordinate $x = (1-r^2)/(1+r^2)$, is suited to extended,
self-similar profiles such as the dilation modes discussed in
\citet{PS23,PS24}. Its elements decay as $\mathcal{O}(1/r^2)$ in
potential and $\mathcal{O}(1/r^4)$ in density while retaining
strict diagonal biorthogonality, so the Gram matrix is the
identity without any recombination.
\end{enumerate}

Numerical tests based on the local asymptotic error function
$F_A(r)$ confirm that both formulations avoid the slow convergence of standard expansion sets.
The modified Clutton-Brock set converges exponentially, with a
numerical rate $\sigma \approx 1.099$, for the mass-conserving
Plummer-difference potential. The Even-Power basis likewise
converges exponentially for smooth potentials with only even
inverse powers of $r$, at a decrement $\sigma(b) = \ln[(b+1)/(b-1)]$ that is
tunable through the core scale $b$; only for modes whose
asymptotic tail contains odd inverse powers of $r$, such as the
isochrone dilation mode, does the rate reduce to an algebraic law
$\propto 1/A$.

Building the correct behaviour at spatial infinity into the basis
functions reduces the size of the Kalnajs response matrix required
for a given accuracy. Because the mass-conservation constraint is
embedded in the expansion itself, rather than imposed as an
external numerical condition, both bases are well suited to
studies of the damping of radial perturbations in open stellar
systems, in particular of the possible existence of damped Landau
quasi-modes~\citep[see, e.g.,][]{Wei_94,PSB}.

Finally, we note that the modified Clutton-Brock and Even-Power bases
constructed in this paper have linear spans consisting, for large $r$,
of only odd and only even powers of $1/r$, respectively. We have
recently constructed a new basis, more suitable for the matrix method,
whose asymptotic tail begins with $1/r^2$ and includes all powers of
$1/r$, the detailed analysis of which is currently in preparation.

\section*{Acknowledgements}
We thank Prasenjit Saha for the constructive question that led to the discussion of non-sparse Gram matrices in Section~2.1. This research was partially supported by the Russian Academy of Sciences Program No. 28 (subprogram II, `Astrophysical Objects as Cosmic Laboratories') and the Ministry of Science and Higher Education of the Russian Federation (I. Shukhman).

\section*{Data Availability}
No new observational data were generated or analysed in support of this research. The \textsc{python} scripts used to produce the figures will be shared on reasonable request to the corresponding author.

\bibliographystyle{mnras}
\bibliography{Clutton_Brock}

\bsp    % typesetting comment
\label{lastpage}
\end{document}